\documentclass[aps,prl,twocolumn,10pt,amsmath,superscriptaddress,amssymb,longbibliography]{revtex4-2}
\usepackage[normalem]{ulem}
\usepackage{epsfig}
\usepackage{epstopdf}
\usepackage{dcolumn}
\usepackage{amsbsy}
\usepackage{bm}
\usepackage{color,CJK}
\usepackage{amsmath,amssymb,amsfonts}
\usepackage{appendix}
\usepackage{graphicx}
\usepackage{algorithm}
\usepackage{enumerate}
\usepackage{makeidx}
\usepackage{braket}
\usepackage[usenames,dvipsnames]{xcolor}
\usepackage[driverfallback=dvipdfmx,colorlinks,linkcolor=blue,citecolor=blue,urlcolor=blue]{hyperref}

\usepackage{pifont}
\usepackage{graphicx}
\usepackage{textcomp}
\usepackage{helvet}

\begin{document}

\title{Experimental study of magnetically insensitive transitions in ultracold Fermi gas of $^{40}$K}

\author{Biao Shan}
\thanks{These authors contributed equally to this work.}
\affiliation{State Key Laboratory of Quantum Optics Technologies and Devices, \\  Institute of Opto-electronics, Collaborative Innovation Center of Extreme Optics, Shanxi University, Taiyuan, Shanxi 030006, China}

\author{Lianghui Huang}
\thanks{These authors contributed equally to this work}
\email[Contact author:]{huanglh06@sxu.edu.cn;}
\affiliation{State Key Laboratory of Quantum Optics Technologies and Devices, \\  Institute of Opto-electronics, Collaborative Innovation Center of Extreme Optics, Shanxi University, Taiyuan, Shanxi 030006, China}
\affiliation{Hefei National Laboratory, Hefei, Anhui 230088, China.}

\author{Yajing Yang}
\affiliation{State Key Laboratory of Quantum Optics Technologies and Devices, \\  Institute of Opto-electronics, Collaborative Innovation Center of Extreme Optics, Shanxi University, Taiyuan, Shanxi 030006, China}

\author{Yuhang Zhao}
\affiliation{State Key Laboratory of Quantum Optics Technologies and Devices, \\  Institute of Opto-electronics, Collaborative Innovation Center of Extreme Optics, Shanxi University, Taiyuan, Shanxi 030006, China}

\author{Jiahui Shen}
\affiliation{State Key Laboratory of Quantum Optics Technologies and Devices, \\  Institute of Opto-electronics, Collaborative Innovation Center of Extreme Optics, Shanxi University, Taiyuan, Shanxi 030006, China}

\author{Zhuxiong Ye}
\affiliation{State Key Laboratory of Quantum Optics Technologies and Devices, \\  Institute of Opto-electronics, Collaborative Innovation Center of Extreme Optics, Shanxi University, Taiyuan, Shanxi 030006, China}

\author{Liangchao Chen}
\affiliation{State Key Laboratory of Quantum Optics Technologies and Devices, \\  Institute of Opto-electronics, Collaborative Innovation Center of Extreme Optics, Shanxi University, Taiyuan, Shanxi 030006, China}
\affiliation{Hefei National Laboratory, Hefei, Anhui 230088, China.}

\author{Zengming Meng}
\affiliation{State Key Laboratory of Quantum Optics Technologies and Devices, \\  Institute of Opto-electronics, Collaborative Innovation Center of Extreme Optics, Shanxi University, Taiyuan, Shanxi 030006, China}
\affiliation{Hefei National Laboratory, Hefei, Anhui 230088, China.}

\author{Pengjun Wang}
\affiliation{State Key Laboratory of Quantum Optics Technologies and Devices, \\  Institute of Opto-electronics, Collaborative Innovation Center of Extreme Optics, Shanxi University, Taiyuan, Shanxi 030006, China}
\affiliation{Hefei National Laboratory, Hefei, Anhui 230088, China.}

\author{Wei Han}
\affiliation{State Key Laboratory of Quantum Optics Technologies and Devices, \\  Institute of Opto-electronics, Collaborative Innovation Center of Extreme Optics, Shanxi University, Taiyuan, Shanxi 030006, China}
\affiliation{Hefei National Laboratory, Hefei, Anhui 230088, China.}

\author{Jing Zhang}
\email[Contact author:]{jzhang74@sxu.edu.cn}
\affiliation{State Key Laboratory of Quantum Optics Technologies and Devices, \\  Institute of Opto-electronics, Collaborative Innovation Center of Extreme Optics, Shanxi University, Taiyuan, Shanxi 030006, China}
\affiliation{Hefei National Laboratory, Hefei, Anhui 230088, China.}

\date{\today }

\begin{abstract}
This paper presents an experimental study of microwave single-photon transitions that are magnetic-field-insensitive in degenerate Fermi gases of $^{40}$K.
This contrasts with microwave single-photon clock transitions for 0-0 magnetic-field-insensitive states and two-photon clock transitions for non 0-0 magnetic-field-insensitive states in bosonic alkali metal atoms.
We show that there are two sets of special transitions between two different hyperfine ground states ($|F$=9/2, $m_{F}$=1/2$\rangle$ $\Leftrightarrow$ $|$7/2, -1/2$\rangle$ and $|$9/2, -1/2$\rangle$ $\Leftrightarrow$ $|$7/2, 1/2$\rangle$), whose microwave single-photon transition frequency is insensitive to low magnetic fields, as the first-order Zeeman shift is almost completely canceled.
By using the microwave spectrum and Ramsey interference fringes, we demonstrate the long-time stability of the coherent transition under magnetic field fluctuations.
These magnetic-field-insensitive microwave hyperfine transitions in ultracold $^{40}$K Fermi gases offer promising applications in quantum information and precision measurements.
\end{abstract}

\maketitle

The development of quantum systems with long coherence times is critical for a wide range of applications, from atomic clocks~\cite{Ye2024PRL} to advanced quantum simulation platforms~\cite{Peter2010NP}. Magnetically insensitive states correspond to pairs of Zeeman sublevels whose transition frequency remains nearly constant under weak fluctuations of the external magnetic field, even though the individual energy levels shift with the field. Coherence between such states can persist for several seconds, enabling their widespread use in high-precision atomic clocks and quantum information applications. In particular, the stabilization of the 0–0 transition line in cold atoms~\cite{AndreasBauch2003IOP,ChristophAffolderbach2005AIP} has driven significant advances in precision measurement technologies, providing a highly reliable time standard~\cite{Golovizin2019NC,Andrew2018PRA,LiuPeng2015PRA}. Furthermore, in arrays of neutral atoms, magnetically insensitive states can serve as robust qubit encodings, enabling long-lived quantum information storage~\cite{Lukin2022Nat}, high-fidelity parallel entangling gates~\cite{Lukin2023Nature1}, and scalable logical quantum processors~\cite{Lukin2023Nature2}.

Bosonic alkali atoms with half-integer nuclear spins possess $m_{F}$=0 Zeeman states, which eliminate the first order Zeeman shift at low magnetic fields.
Extensive theoretical and experimental work has focused on 0--0 microwave (MW) single-photon transitions between the $m_F = 0$ Zeeman sublevels of bosonic atoms, including $^{23}\mathrm{Na}$ ($I = 3/2$)~\cite{Ketterle2003PRL,Atsuo2010PRA}, $^{39}\mathrm{K}$ ($I = 3/2$)~\cite{Antoni2017PRA}, $^{87}\mathrm{Rb}$ ($I = 3/2$)~\cite{Zibrov2010PRA,Lukin2022PRA}, and $^{133}\mathrm{Cs}$ ($I = 7/2$)~\cite{Chu1993PRL,Salomon1999PRL}.
Furthermore, non 0--0 magnetically insensitive transitions in bosonic alkali-metal atoms can also be realized by eliminating the first-order differential Zeeman shift (i.e., the difference in Zeeman shifts between the two states) at low magnetic fields, such as the $\lvert F=1, m_F=\mp 1\rangle \Leftrightarrow \lvert F=2, m_F=\pm 1\rangle$ transition of $^{87}$Rb~\cite{Lin2013JPB,Bresson2022PRA}.
However, these transitions generally require two-photon or Raman driving.
In addition, non 0--0 magnetically insensitive transitions can also occur in bosonic alkali-metal atoms at specific magnetic fields where the first- and higher-order differential Zeeman shifts cancel each other out~\cite{Cornell2002PRA,Marcassa2006Sciene, Philipp2009NP, Porto2011PRL, Weiss2014PRA,Shougang2020PhysScr}.

There have been few theoretical or experimental reports on magnetically insensitive transitions for fermionic alkali atoms with integer nuclear spins, primarily due to the absence of $m_F = 0$ Zeeman states.
Only recently have non-0–0 magnetically insensitive transitions been experimentally realized in fermionic $^{6}$Li using two-photon Raman processes, and they have been successfully employed to implement cold-atom interferometers and to precisely measure recoil frequencies~\cite{Haibin2023PRR}.
In this paper, we experimentally demonstrate that non-0–0 magnetically insensitive transitions can also be realized via MW single-photon processes in $^{40}$K atoms, enabling the near-complete cancellation of the first-order differential Zeeman shift.
The stability of these transitions is investigated using MW spectroscopy and Ramsey interferometry, showing that magnetically insensitive transitions allow for narrow-linewidth spectra and precise measurements of hyperfine frequency shifts.

\begin{figure}[tb]
\centering
\includegraphics[width=3.0in]{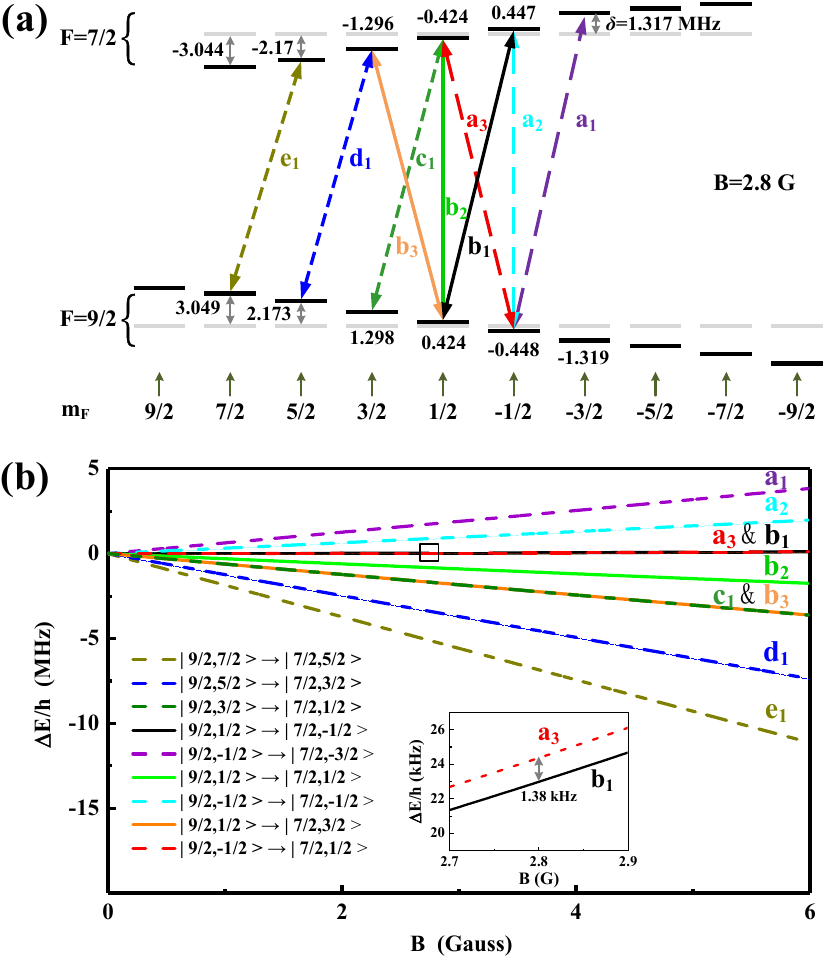}
\caption{(a)~Scheme of the energy levels and transitions among the lowest hyperfine Zeeman states of $^{40}$K. The gray lines indicate the energy levels at zero magnetic field, while the black lines represent the Zeeman-split energy levels at 2.8 G. Nine sets of microwave transitions between the Zeeman sublevels are considerred in our current study, involving $a_{i}$, $b_{i}$, $c_{i}$, $d_{i}$, and $e_{i}$ transitions, where $i$=1, 2, 3 corresponds to $\sigma^{+}, \pi, \sigma^{-}$ transitions.
(b)~Differential Zeeman shifts as a function of magnetic field. Two magnetically insensitive transitions, $a_{3}$ and $b_{1}$, are identified, which exhibit minimal dependence on the magnetic field. The inset shows an enlarged view of the differential Zeeman shifts for the $a_{3}$ and $b_{1}$ transitons, with the transition frequency difference being approximately 1.38 kHz near a magnetic field of 2.8 G.
}
\label{energy}
\end{figure}

The hyperfine energy levels in the ground state manifold  of the $^{40}$K atom are shown in Fig.~\ref{energy}. When subjected to a magnetic field \( B \), the Zeeman shifts of the hyperfine states depend on the magnetic quantum number $m_{F}$ and can be described by the $Breit$-$Rabi$ $formula$~\cite{Rabi1931PR,Daniel2019K40}
\begin{equation}\label{Breit}
\begin{split}
E_{B}=&-\frac{A}{4}+\mu_B g_{I}m_{F} B \\ & \pm\frac{A(2I+1)}{4}\sqrt{1+\frac{4m_{F}}{2I+1}x+x^{2}},
\end{split}
\end{equation}
where $A=-h\times 285.7308$ MHz is hyperfine interaction constant~\cite{Daniel2019K40,Davis1949PR, Eisinger1952PR, Arimondo1977RMP, Allegrini2022JPC}, $x$$=$$\frac{2(g_J-g_I)\mu_B}{A(2I+1)} B$, $\mu_B$ is the Bohr-magneton, $g_J$ and $g_I$ are the Land\'{e} g-factor of the electron and the nucleus, respectively, with $g_J$/$g_I$$\approx$10$^{4}$~\cite{Daniel2019K40}.
In the low-magnetic-field regime, Eq.~(\ref{Breit}) reduces to the first-order form
\begin{eqnarray}\label{jinsi}
\begin{aligned}
E^{+}_{B} &\approx-\frac{A}{4}+\frac{\Delta E_{hf}}{2}+\frac{\mu_{B}g_{J}}{\left(  2I+1\right)  }m_{F}B
& \\
E^{-}_{B} &\approx-\frac{A}{4}-\frac{\Delta E_{hf}}{2}-\frac{\mu_{B}g_{J}}{\left(  2I+1\right)  }m_{F}B
&
\end{aligned}
\end{eqnarray}
where $E^{+}_{B}$ and $E^{-}_{B}$  are the hyperfine energy levels for the $ F = 9/2$  and $ F = 7/2$ manifolds, respectively, and $\Delta E_{hf}=A(I+1/2)=-h \times 1285.7886$ MHz is the zero-field hyperfine split of the ground state for $^{40}$K atoms\cite{Daniel2019K40,Davis1949PR, Eisinger1952PR, Arimondo1977RMP, Allegrini2022JPC}.

From Eq.~(\ref{jinsi}), one can find that the energy $E^{-}_{B}$ of the Zeeman sublevels for the $F = 7/2$ manifold increases linearly as the magnetic quantum number $m_F$ decreases, while the energy $E^{+}_{B}$ for the $F = 9/2$ manifold decreases correspondingly, as shown in Fig.~\ref{energy}(a).
In particular, there exist two sets of magnetically insensitive MW single-photon transitions ($a_3$: $|9/2, -1/2\rangle \Leftrightarrow |7/2, 1/2\rangle$ and $b_1$: $|9/2, 1/2\rangle \Leftrightarrow |7/2, -1/2\rangle$), for which the first-order differential Zeeman shifts, defined as $\Delta E(B) = E^{-}_{B} - E^{+}_{B} - \Delta E_{\mathrm{hf}}$, are nearly completely canceled at low magnetic fields, as shown in Fig.~\ref{energy}(b).
From Fig.~\ref{energy}(b), among the nine considered single-photon transitions ($a_{1}$, $a_{2}$, $a_{3}$, $b_{1}$, $b_{2}$, $b_{3}$, $c_{1}$, $d_{1}$, and $e_{1}$), only $a_3$ and $b_1$ are magnetically insensitive, exhibiting an almost negligible response to the magnetic field with $\delta(\Delta E)/\delta B \approx 0 $. In contrast, the other transitions show significant frequency shifts as the magnetic field varies.

\begin{figure}[tb]
\centering
\includegraphics[width=2.6in]{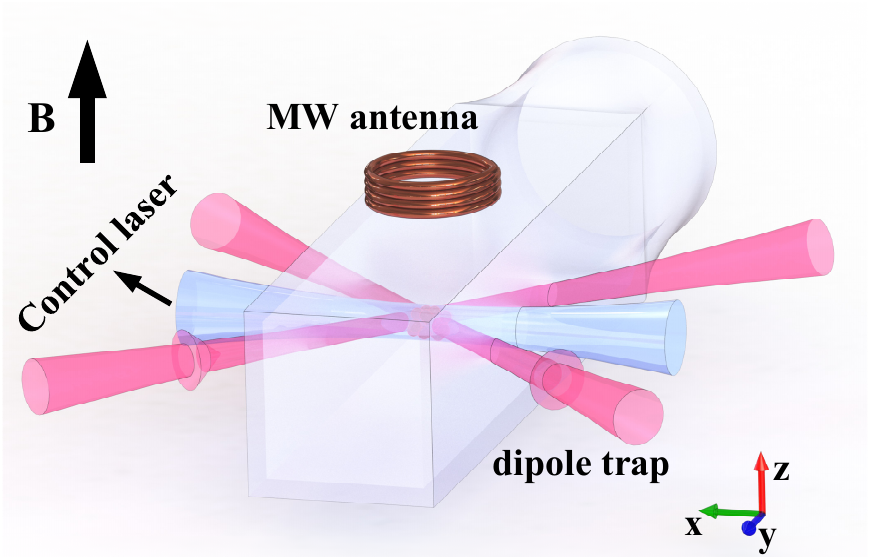}
\caption{Experimental setup. The atoms are confined in a crossed optical dipole trap formed by 1064-nm lasers. The control laser for the ac-Stark shift propagates along the $\hat{x}$ direction and is collimated with a Gaussian beam waist of 2 mm. A magnetic field applied along the $\hat{z}$ direction induces Zeeman splitting and serves as the quantization axis. A microwave antenna generates a signal along the $\hat{z}$ direction to drive MW transitions.
}
\label{setup}
\end{figure}

\begin{figure}[tb]
\includegraphics[width=3.4in]{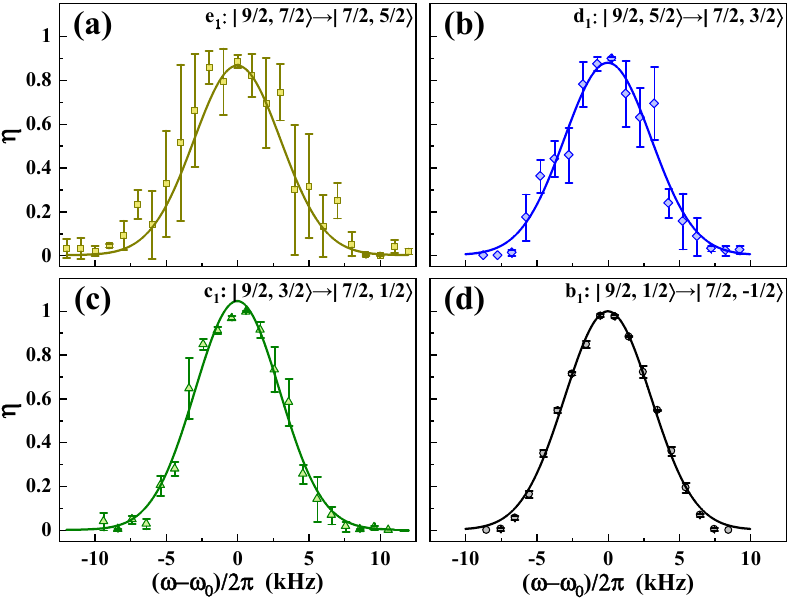}
\caption{Microwave spectra of the hyperfine transitions: (a) $e_1$, (b) $d_1$, (c) $c_1$, and (d) $b_1$. The measurements were performed using a Gaussian-shaped $\pi$ pulse with a duration of 0.11~ms and an optimized amplitude under a bias magnetic field of 2.8~G. The spectra are quantified by the population ratio $\eta = N_2 / (N_1 + N_2)$, where $N_1$ and $N_2$ denote the atom numbers of the hyperfine Zeeman states for each transition. The full width at half maximum (FWHM) of the observed spectra is $2\pi \times \{7.18, 7.18, 7.02, 7.14\}$~kHz for the respective transitions, extracted by fitting the experimental data with the Gaussian pulse formula $\eta_1(\delta) = \eta_0 \exp\left(-\delta^2 / 2\sigma^2\right)$. The center transition frequency $\omega_0$ is determined by $\omega_0 = \Delta E_{hf}/\hbar + \Delta E(B)/\hbar$, where $\Delta E(B)/\hbar$ is the magnetic-field-induced frequency shift, with $\Delta E(B)/\hbar = 2\pi \times (-5.219)$~MHz for $e_1$, $2\pi \times (-3.469)$~MHz for $d_1$, $2\pi \times (-1.722)$~MHz for $c_1$, and $2\pi \times (0.023)$~MHz for $b_1$. The symbol points represent experimental data, and the solid lines indicate the Gaussian fit of the data. Error bars represent the standard deviation from five repeated measurements.}
\label{clock}
\end{figure}

\begin{figure}[tb]
\includegraphics[width=2.7in]{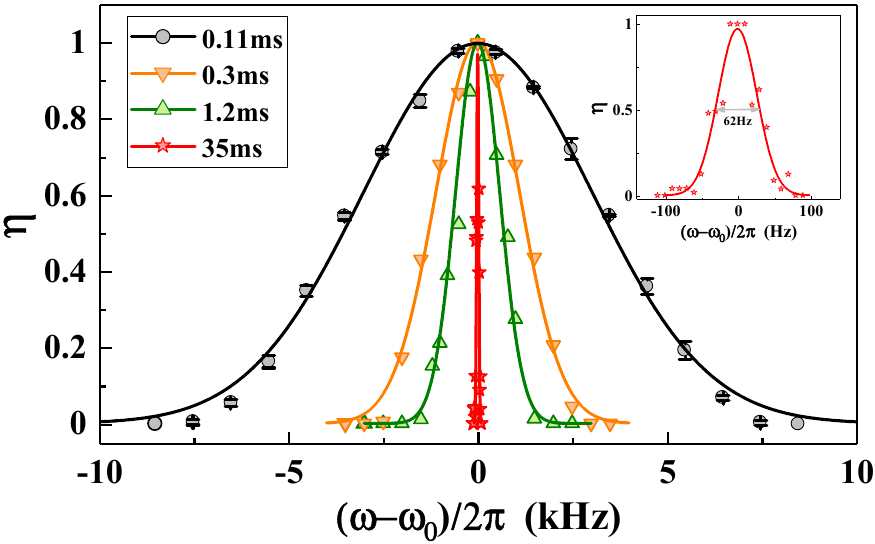}
\caption{Microwave spectra of the hyperfine transition $b_1$ for different durations of Gaussian $\pi$ pulses. The spectra correspond to pulse durations of 0.11~ms (black circles), 0.3~ms (orange inverted triangles), 1.2~ms (green triangles), and 35~ms (red pentagons), with respective FWHMs of $2\pi \times \{7140, 2600, 1380, 62\}$~Hz. The inset highlights the spectrum for the 35~ms pulse, showing a FWHM of $2\pi \times 62$~Hz. Different marker shapes represent experimental data, while the colored lines indicate Gaussian fits to the data.
\label{narrow}}
\end{figure}

A detailed description of the experimental setup is available in Refs.~\cite{WeiDong2007CPL,chaiShijie2012ActaSinQuantumOpt,Weidong2006ActaPhysSin,Wangaqiong2016ASQO,liangchao2017ASQO,MiaoJie2022CPB,Ding2024CPB}.
As illustrated in Fig.~\ref{setup}, a degenerate Fermi gas of $^{40}$K is captured in a crossed optical dipole trap. Approximately $N=4\times10^{6}$ ultracold $^{40}$K atoms are initially prepared in the hyperfine Zeeman state $|9/2, 9/2\rangle$  at a temperature of $0.3T_F$, where the Fermi temperature is defined as $T_F=\hbar \bar{\omega} (6N)^{1/3}/k_B$.
Here, $\bar{\omega}=(\omega_x\omega_y\omega_z)^{1/3}\approx 2\pi\times80$ Hz is the geometric mean trapping frequency, $N$ is the total atom number, and $k_B$ is the Boltzmann constant. Using the Landau--Zener transition, we adiabatically prepare atoms in arbitrary Zeeman states starting from the initial state \(|9/2,9/2\rangle\). This preparation is realized by sweeping a radio-frequency (RF) field over 100~ms under a magnetic field of 19.6~G. The RF frequency ramp starts at 6.56~MHz, and its end frequency determines the target Zeeman state, for example 6.30~MHz for \(|9/2,7/2\rangle\), 6.24~MHz for \(|9/2,5/2\rangle\), 6.18~MHz for \(|9/2,3/2\rangle\), and 6.12~MHz for \(|9/2,1/2\rangle\)~\cite{KlemptPhD2007}.

Then we drive the transition of $^{40}$K atoms from an arbitrary Zeeman state in the $F=9/2$ manifold to a final Zeeman state in the $F=7/2$ manifold at a low external magnetic field of 2.8 Gauss. To evaluate the stability of this transition, we employ a Gaussian-shaped MW pulse~\cite{Pert2002OC},which can avoid the appearance of sidelobes in the MW spectrum~\cite{Pengjun2012PRA,Zhengkun2012PRA}. The MW pulse is generated by mixing MW and RF signals, which consists of a main signal 1.286 GHz and a Gaussian-shaped RF sideband signal around 50 MHz~\cite{Ziliang2023CPB}. Then it is amplified to a maximum output power of 5 W by an amplifier (Mini-Circuits, ZHL-5W-2G-S+), and applied to the atoms in the vertical direction through a MW antenna, as shown in Fig.~\ref{setup}. The coupling strength of atomic transition is controlled by adjusting the amplitude of the Gaussian sideband signal. To measure the populations of the initial and final states, we again employ an adiabatic Landau–Zener transition to transfer the atoms from the $F=7/2$ manifold back into the $F=9/2$ manifold, but into a Zeeman state distinct from the original $F=9/2$ state. With the aid of a Stern–Gerlach gradient magnetic field, the populations originating from both the initial $F=9/2$ state and the final $F=7/2$ state can then be simultaneously resolved by time-of-flight (TOF) absorption imaging.

By fixing the width of the Gaussian pulse with a duration of 0.11~ms, with an appropriate amplitude corresponding to a $\pi$ pulse at resonance, we scan the frequency to measure the microwave spectrum for the hyperfine transitions $e_1: |9/2, 7/2\rangle \Leftrightarrow |7/2, 5/2\rangle$, $d_1: |9/2, 5/2\rangle \Leftrightarrow |7/2, 3/2\rangle$, $c_1: |9/2, 3/2\rangle \Leftrightarrow |7/2, 1/2\rangle$, and $b_1: |9/2, 1/2\rangle \Leftrightarrow |7/2, -1/2\rangle$, as shown in Fig.~\ref{clock}. To verify the stability of magnetically insensitive transitions, we evaluated the error bars of the transition spectra based on five repeated measurements. The microwave spectra in Figs.~\ref{clock}(a)-(b) for the transitions $e_1$$:\ |9/2, 5/2\rangle \Leftrightarrow |7/2, 3/2\rangle$ and $d_1$$:\ |9/2, 7/2\rangle \Leftrightarrow |7/2, 5/2\rangle$ exhibit large errors, reflecting significant frequency shifts induced by magnetic field fluctuations. In contrast, the spectrum in Fig.~\ref{clock}(c) for the transition $c_1$$:\ |9/2, 3/2\rangle \Leftrightarrow |7/2, 1/2\rangle$ displays only minor fluctuations, indicating a relatively small frequency shift. Notably, the spectrum in Fig.~\ref{clock}(d) for the transition $b_1$:$\ |9/2, 1/2\rangle \Leftrightarrow |7/2, -1/2\rangle$ exhibits minimal errors and demonstrates excellent stability against magnetic field fluctuations, thereby verifying the magnetic insensitivity of this transition.

\begin{figure}[!htb]
\includegraphics[width=2.6in]{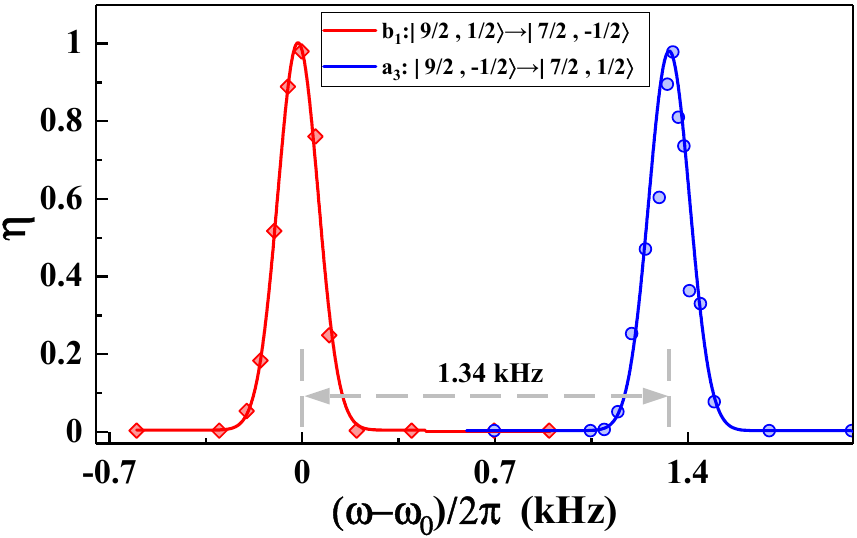}
\caption{Microwave spectra of two sets of magnetically insensitive transitions, $b_1$ and $a_3$. The measurements are performed using a Gaussian-shaped $\pi$ pulse with a duration of 10.5 ms and optimized amplitude under a bias magnetic field of 2.8 G. The FWHMs for transitions $b_1$ and $a_3$ are $2\pi \times$172 Hz and $2\pi \times$175 Hz, respectively. The measured transition frequency difference between the two magnetically insensitive transitions is 1.34~kHz, which is consistent with the theoretical calculation shown in the inset of Fig.~1(b). Different shaped markers represent the experimental data, while colored lines indicate the results of Gaussian fits.
\label{scanning}}
\end{figure}

Taking advantage of the magnetic insensitivity of the $b_1$ transition, we narrow the microwave spectrum by increasing the duration of the Gaussian pulse, enabling precise characterization of this transition, as shown in Fig.~\ref{narrow}. By extending the Gaussian pulse duration from 0.12~ms to 35~ms, the FWHM decreases from $2\pi \times 7.14$~kHz to $2\pi \times 62$~Hz. In addition, there also exists another magnetically insensitive transition, $a_3$: $\ |9/2, -1/2\rangle \Leftrightarrow |7/2, 1/2\rangle$, as shown in Fig.~1(b), which lies only 1.3~kHz away from the $b_1$ transition. The ability to achieve such a narrow linewidth, enabled by magnetic insensitivity, allows us to clearly resolve these two closely spaced transitions. Fig.~\ref{scanning} shows the microwave spectrum for both transitions with high distinguishability.

\begin{figure}[tb]
\includegraphics[width=3.42in]{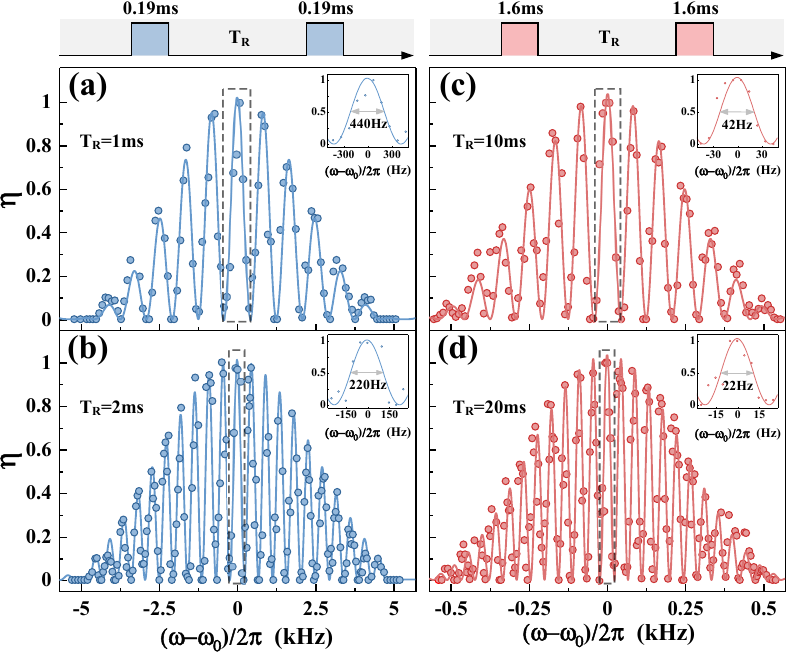}
\caption{
Ramsey fringe spectra for the magnetically insensitive $b_1$ transition.
The upper panels show the Ramsey timing sequence.
Circles represent the experimental data, and lines show fits using Eq.~(\ref{3}).
Pulse durations are $\tau_p = 0.19$~ms ($\pi/2$ pulses) for panels (a) and (b), and $\tau_p = 1.6$~ms ($\pi/2$ pulses) for panels (c) and (d).
Insets display the central Ramsey fringe with FWHM $\simeq 2\pi \times (440, 220, 42, 22~\textrm{Hz})$.
\label{Ramsey}}
\end{figure}

Now, we focus on the Ramsey interference for the magnetically insensitive transition $b_1$: $\ |9/2, 1/2\rangle \Leftrightarrow |7/2, -1/2\rangle$. Ramsey interference employs two short pulses separated by a free evolution time to probe small deviations between the atomic transition frequency and the driving frequency by monitoring changes in the atomic populations. The pulse sequence is illustrated in the upper panels of Fig.~\ref{Ramsey}. In our experiment, we apply two identical square pulses with a fixed duration and an appropriate amplitude corresponding to a resonant $\pi/2$ pulse. For a free evolution time $T_R$, the Ramsey fringes are obtained by measuring the populations $N_{1}$ and $N_{2}$ in the initial and final states while scanning the detuning $\delta$ from resonance. The spectral envelope of the Ramsey interference pattern is determined by the pulse duration $\tau_p$, consistent with the MW spectrum discussed above, with shorter pulses leading to broader envelopes~\cite{W.Vassen2016APB}. For a given $\tau_p$, the number of observable fringes increases with the free evolution time $T_R$, which thus determines the frequency resolution.

\begin{figure}[tb]
\includegraphics[width=2.6in]{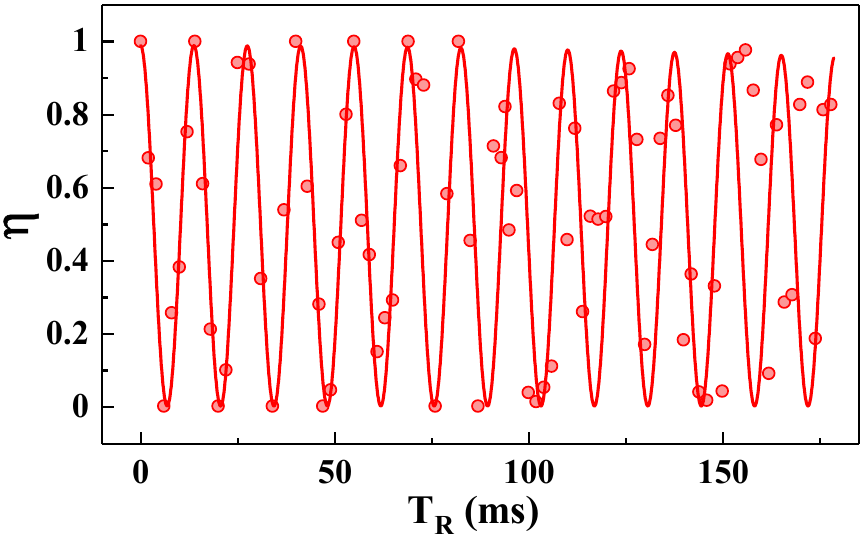}
\caption{
Ramsey oscillations as a function of the free evolution time $T_R$ for the $b_1$ transition, measured using a $\pi/2$ pulse of duration $\tau_p = 0.19$ ms.
With the MW frequency fixed at $\omega = 2\pi \times 1285.811551~\textrm{MHz}$, the Ramsey detuning is determined to be $\delta / 2\pi \approx 73 \pm 5~\textrm{Hz}$, yielding a resonance frequency of $\omega_0 = 2\pi \times 1285.811624(5)~\textrm{MHz}$ under a bias magnetic field of 2.8~G.
Red points represent the experimental data, and the solid red line is a fit using Eq.~(\ref{3}).
\label{Oscillation}}
\end{figure}

\begin{figure*}[tb]
\includegraphics[width=5.0in]{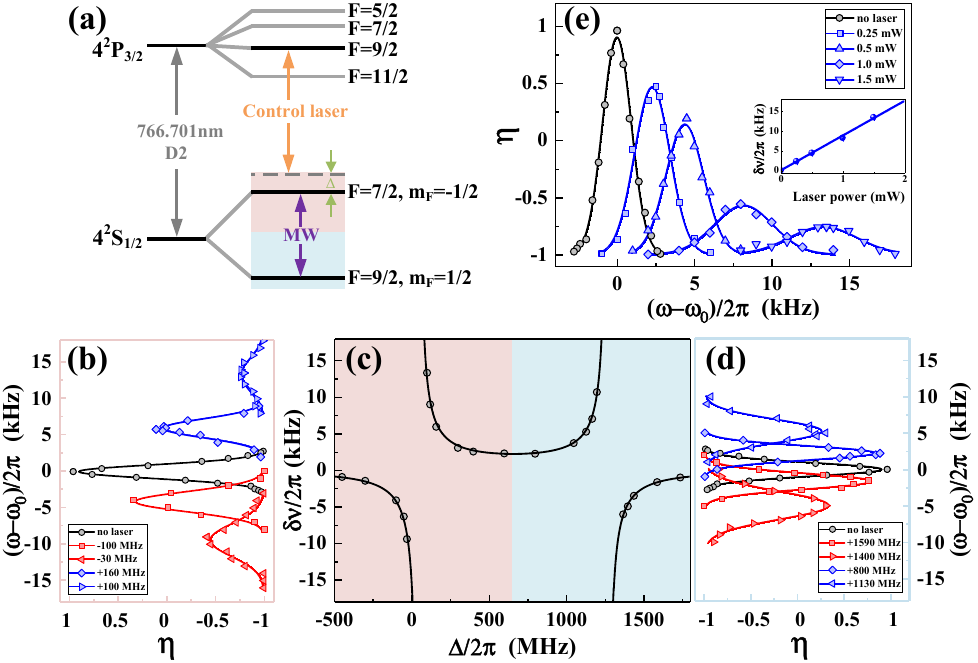}
\caption{The differential frequency shift $\delta\nu$ between two magnetically insensitive states controlled by a detuned laser. (a)~The relevant energy levels and transitions for $^{40}$K atoms. While the transition frequency from the Zeeman state $|4S_{1/2}, F = 7/2, m_F=-1/2\rangle$ to the hyperfine manifold $|4P_{3/2}, F = 9/2\rangle$ serves as a reference for calibrating the control laser detuning, the microwave transition $b_1$  between the two magnetically insensitive states $|4S_{1/2}, F = 9/2, m_F=1/2\rangle$ and $|4S_{1/2}, F = 7/2, m_F=-1/2\rangle$ is used to measure the differential frequency shift $\delta\nu$. (b), (d)  Microwave spectra measured for various laser detunings near the resonance of the $|4S_{1/2}, F = 7/2, m_F = -1/2\rangle \Leftrightarrow |4P_{3/2}, F = 9/2\rangle$ transition, corresponding to the red-shaded and blue-shaded regions in (c), respectively. (c) The differential frequency shift $\delta\nu$ between the two magnetically insensitive states for the $b_1$ microwave transition, plotted as a function of the control laser detuning $\Delta$, which is obtained from the data of (b) and (d). The laser power is fixed at 1.5 mW in (b), (c) and (d).
(e) Microwave spectra for different laser powers at a fixed laser detuning of $\Delta/2\pi = +100~\mathrm{MHz}$. The inset of (e) shows the relationship between laser power and the differential frequency shift $\delta\nu$.
\label{shift}}
\end{figure*}

The experimental Ramsey interference fringes can be fitted with the following expression~\cite{AndreaAlberti2021PRApple}:
\begin{equation}\label{3}
\eta_{2} = \left|\frac{\tilde{\Omega}\tau_p}{2}\right|^2
\left[\frac{\sin(\delta \tau_p/2)}{\delta \tau_p/2}\right]^2
\cos^2\!\left(\frac{\delta T_R}{2}\right),
\end{equation}
where $\tilde{\Omega} = \sqrt{\Omega^2 + \delta^2}$ is the effective coupling strength of the MW field, $\tau_p$ is the duration of the $\pi/2$ square pulse, $\delta = \omega - \omega_0$ is the detuning of the MW field from the magnetically insensitive transition, and $T_R$ is the free evolution time. Two sets of typical Ramsey fringes with $T_R = 1~\textrm{ms}$ and $T_R = 2~\textrm{ms}$, obtained using a $\pi/2$ pulse duration of $\tau_p = 0.19~\textrm{ms}$, are shown in Figs.~\ref{Ramsey}(a) and (b). The FWHM of the central peak is $2\pi \times 440~\textrm{Hz}$ for $T_R = 1~\textrm{ms}$ and $2\pi \times 220~\textrm{Hz}$ for $T_R = 2~\textrm{ms}$. By increasing both the pulse width $\tau_p$ and the free evolution time $T_R$, the Ramsey spectrum exhibits a narrower envelope (set by $\tau_p$) and a reduced central fringe linewidth (set by $T_R$), as shown in Figs.~\ref{Ramsey}(c) and (d). These correspond to $T_R = 10~\textrm{ms}$ and $T_R = 20~\textrm{ms}$ with $\tau_p = 1.6~\textrm{ms}$ pulses. The FWHM of the central fringe decreases from $2\pi \times 42~\textrm{Hz}$ at $T_R = 10~\textrm{ms}$ to $2\pi \times 22~\textrm{Hz}$ at $T_R = 20~\textrm{ms}$. The well-resolved and stable Ramsey fringes not only demonstrate the robustness of the magnetically insensitive transition but also highlight its potential as a highly sensitive spectroscopic probe~\cite{Gallagher1992PRL}.

We can precisely determine the central transition frequency $\omega_0$ by measuring the Ramsey oscillations as a function of the free evolution time $T_R$.
Setting the MW frequency to an arbitrary value near resonance with $\omega =2\pi \times 1285.811551~\textrm{MHz}$, we record the Ramsey oscillations while varying $T_R$, as shown in Fig.~\ref{Oscillation}.
As the period of the Ramsey oscillations is inversely related to the detuning, $T_\mathrm{osc} = 2\pi / \delta$, we extract a Ramsey detuning of $\delta / 2\pi \approx 73 \pm 5~\textrm{Hz}$ from the measured oscillation curves, yielding a resonance frequency of $\omega_0 = 2\pi \times 1285.811624(5)~\textrm{MHz}$ for the $b_1$: $\ |9/2, 1/2\rangle \Leftrightarrow |7/2, -1/2\rangle$ transition under a bias magnetic field of 2.8~G.

\begin{figure*}[!htb]
\centering
\includegraphics[width=5.2in]{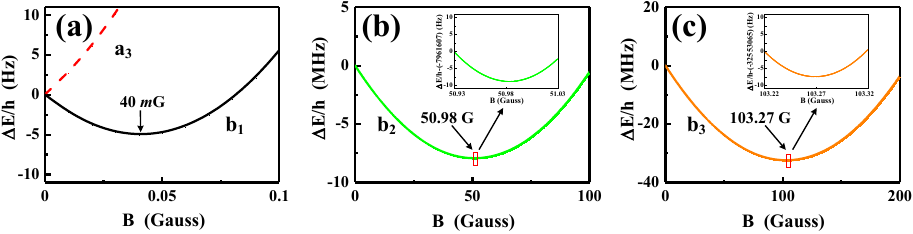}
\caption{Magnetically insensitive transitions with complete Zeeman shift cancellation near specific magnetic fields. Three magnetically insensitive transitions, $b_1$, $b_2$, and $b_3$, are identified, each exhibiting zero magnetic field dependence near 40~mG, 50.98~G, and 103.27~G, respectively. The insets show zoomed-in views around the minima in panels (b) and (c).
}
\label{point}
\end{figure*}

Furthermore, we examine the influence of a detuned laser on the magnetically insensitive transitions. The interaction between the laser and the atoms induces an ac-Stark shift, which modifies the atomic energy levels. The ac-Stark shift is generally determined by the scalar, vector, and tensor polarizabilities of the atoms. The scalar polarizability induces a uniform energy shift that is independent of the Zeeman state but differs between different hyperfine levels. The vector polarizability causes a shift that depends on the magnetic quantum number \(m_F\), with the magnitude of the shift varying linearly with \(m_F\). This contribution is negligible when the laser propagates perpendicular to the magnetic quantization axis~\cite{Mitroy2013PRA, Widera2016PRA, Wen2021JOSAB}, which is the case in our experiment, as shown in Fig.~2. Additionally, the tensor polarizability is weak in our current experiment. Consequently, only the scalar light shift contributes significantly, and the effects of laser polarization can be ignored.

The differential shift \(\delta \nu = \delta \nu_1 - \delta \nu_2\) between the magnetically insensitive states \( |9/2, 1/2\rangle \) and \( |7/2, -1/2\rangle \), with \(\delta \nu_1\) and \(\delta \nu_2\) representing the ac-Stark shifts for each state, is shown in Fig.~\ref{shift}(c). When the laser is red-detuned for both magnetically insensitive ground states (\( |9/2, 1/2\rangle \) and \( |7/2, -1/2\rangle \)), the ac Stark shifts of both states are negative, with $\delta\nu_1 < 0$ and $\delta\nu_2 < 0$. Since the detuning of the \( |9/2, 1/2\rangle \) state is larger in magnitude than that of the \( |7/2, -1/2\rangle \) state, as shown in Fig.~\ref{shift}(a), we have $|\delta\nu_1| > |\delta\nu_2|$, resulting in a differential shift $\delta\nu = \delta\nu_1 - \delta\nu_2 < 0$. When the laser is blue-detuned for both the \( |9/2, 1/2\rangle \) and \( |7/2, -1/2\rangle \) states, the ac-Stark shifts are positive, with $\delta\nu_1 > 0$ and $\delta\nu_2 > 0$. However, the detuning of the \( |9/2, 1/2\rangle \) state is smaller, so $|\delta\nu_1| < |\delta\nu_2|$, leading again to a differential shift $\delta\nu = \delta\nu_1 - \delta\nu_2 < 0$. In contrast, when the laser is blue-detuned for the \( |7/2, -1/2\rangle \)) state and red-detuned for the \( |9/2, 1/2\rangle \) state, we have $\delta\nu_1 > 0$ and $\delta\nu_2 < 0$, which yields a positive differential shift $\delta\nu = \delta\nu_1 - \delta\nu_2 > 0$.

The laser-induced differential shift \(\delta\nu\) leads to a corresponding frequency shift in the resonant MW transition between the two magnetically insensitive states. Figs.~\ref{shift}(b) and \ref{shift}(d) show the MW spectra for different laser detunings. It can be found that the central positions of the resonance peaks, which correspond to the resonant MW transition frequencies, shift in accordance with the differential shift \(\delta \nu\).
By fixing the laser detuning, we also investigate the influence of laser power on the differential frequency shift \(\delta \nu\) between the two magnetically insensitive states by measuring the microwave spectra. We find that the measured frequency shift in the resonant MW transition, which corresponds to the laser-induced differential frequency shift \(\delta \nu\), exhibits a clear linear dependence on the laser power, as shown in Fig.~\ref{shift}(e).
In addition, from Figs. ~\ref{shift}(b), ~\ref{shift}(d) and ~\ref{shift}(e), it is found that the spectral line broadens with increasing laser power or decreasing laser detuning. This can be explained as higher laser power or smaller detuning leads to a higher scattering rate~\cite{Metcalf2003JOSAB}, which in turn enhances atomic loss and decoherence.

In conclusion, we have experimentally investigated the microwave single-photon transitions between magnetically insensitive non-0–0 Zeeman states of $^{40}$K atoms.
Specifically, we examined two distinct transitions connecting different hyperfine ground states: $|9/2, 1/2\rangle$$\Leftrightarrow$$|7/2, -1/2\rangle$ and $|9/2, -1/2\rangle$$\Leftrightarrow$$|7/2, 1/2\rangle$.
The spectral properties of these transitions were characterized through microwave spectroscopy and Ramsey interferometry, confirming that these magnetically insensitive transitions support narrow spectral linewidths and enable high-precision determination of hyperfine frequency shifts.
In addition, we also found that near certain specific magnetic fields, some magnetically insensitive transitions can completely eliminate the Zeeman frequency shift, as shown in Fig.~\ref{point}.
These transitions may exhibit enhanced stability against magnetic field fluctuations, which will be the focus of our future experiments.

Magnetic field noise presents a significant challenge in many ultracold atom experiments. For instance, in previous realizations of spin-orbit coupling using Raman laser schemes, the internal atomic states were highly sensitive to magnetic field fluctuations due to Zeeman shifts~\cite{Spielman2011Nat,Fu2011PRA,Wang2012PRL,Huang2016NP}, imposing stringent requirements on magnetic field stability and posing significant challenges for studying spin-orbit coupling effects~\cite{shuai2018SB,shuai2019RSI}.
By utilizing the magnetically insensitive states identified in our current work, the influence of magnetic noise can be effectively suppressed, providing a robust experimental foundation for exploring more complex spin-orbit coupling schemes in ultracold atomic systems ~\cite{Shan2025PRA}.
These magnetically insensitive states can also be applied to extend the coherence time of clock transitions~\cite{Rosenbusch2011PRL}, enhance the measurement accuracy of electromagnetically induced transparency for detecting long-range Rydberg interactions~\cite{Martin2007PRL,Wenhui2016PRA}, and improve the sensitivity of scattering length measurements for probing atomic interactions~\cite{Jin2003PRL,Ketterle2003Science,Ludlow2009Sci}.

Furthermore, the Hz level stability of the transition frequency achieved in the current experiment can be directly applied to measure the ground-state hyperfine structure constant \emph{A} of $^{40}$K, with the potential to improve the measurement precision by 2–3 orders of magnitude compared to conventional results. To date, high-precision determination of this constant in $^{40}$K remains an unmet challenge: its value is currently known only at the kHz level of accuracy~\cite{Daniel2019K40,Davis1949PR, Eisinger1952PR, Arimondo1977RMP, Allegrini2022JPC}. In contrast, the ground-state hyperfine constants of all the other naturally occurring alkali atoms, including $^{6}$Li, $^{7}$Li, $^{23}$Na, $^{39}$K, $^{41}$K, $^{85}$Rb, $^{87}$Rb, and $^{133}$Cs, have been measured with Hz-level or even sub-Hz precision~\cite{Arimondo1977RMP, Allegrini2022JPC}. This disparity arises primarily from the low natural abundance of $^{40}$K, as well as from the limited precision of relevant atomic and spectroscopic parameters. The high-precision detection of magnetically insensitive transitions demonstrated in the present work establishes essential experimental techniques for $^{40}$K-based quantum precision measurements, including—but not limited to—the potential to substantially improve the determination of the hyperfine interaction constant.

\begin{acknowledgments}

This research is supported by National Key Research and Development Program of China (Grants No.~2022YFA1404101, and No.~2021YFA1401700), Innovation Program for Quantum Science and Technology (Grants  No.~2021ZD0302003), National Natural Science Foundation of China (Grants No.~12034011, No.~92476001, No.~U23A6004, {No.~12488301}, No.~12474266, No.~12474252, No.~12374245, No.~12322409, and No.~92576205, and No.~12504336), the Fundamental Research Program of Shanxi Province (No.~202403021221001), and the Fund Program for the Scientific Activities of Selected Returned Overseas Professionals in Shanxi Province(No.~20250004).

\end{acknowledgments}
\bibliography{reference-clock-state}

\end{document}